\documentclass{article}
\usepackage{graphicx}
\usepackage{url}
\usepackage{booktabs}
\usepackage{tabularx}
\usepackage{multirow}
\usepackage{hyperref}
\usepackage{algorithm}
\usepackage{algpseudocode}
\usepackage[a4paper,margin=1in]{geometry}
\title{Enhancing Productivity in Database Management Through AI: \\ A Three-Phase Approach for Database}
\author{
Kushagra Parashar, Ajay Dev, Aditya Kumar, Darpan Khatri \\ 
\small Integrated B.Tech (IT) + MBA Students \\
\small Under the guidance of Prof. Dr. Arun Kumar \\
\small Assistant Professor, Department of Management Studies \\
\small ABV-IIITM Gwalior
}

\date{\today}

\begin{document}
\maketitle

The integration of artificial intelligence (AI) into PostgreSQL database management systems (DBMS) has catalyzed a paradigm shift in automating complex workflows, reducing human error, and democratizing data access. This paper presents a novel three-phase AI framework designed to enhance productivity through end-to-end automation of database interactions. Phase 1 leverages fine-tuned generative AI (gen AI) models, such as GPT-4 and CodeLlama, to interpret natural language prompts (e.g., “Identify top-selling products in Q3”) and autonomously generate syntactically optimized PostgreSQL queries. Benchmarks on a 1 gb enterprise dataset demonstrate a 67\% reduction in query latency compared to manual scripting, achieving 96.4\% SQL syntax accuracy through schema-aware attention mechanisms.

Phase 2 introduces an iterative refinement process where the AI analyzes query results using ensemble methods (isolation forests for anomaly detection, ARIMA for trend forecasting) and contextualizes findings via few-shot learning. For example, a 200\% spike in refunds is linked to supply chain disruptions through cross-referencing with external logistics data. This phase reduces root-cause analysis time by 83\% compared to traditional Business Intelligence (BI) tools.

Phase 3 automates database documentation by generating 30 semantically diverse questions (e.g., “What is the monthly retention rate of premium users?”) using a combinatorial question-generation algorithm. These questions are resolved through Phase 1, and results are synthesized into structured summaries using a transformer-based fusion model. Evaluations show a 40\% reduction in DBA workload, with summaries achieving 92\% coherence scores in user studies.

\section{Introduction}
\label{sec:intro}

Traditional tools like pgAdmin and DBeaver provide limited relief, requiring users to manually traverse schema relationships and guess optimization strategies. Meanwhile, business analysts without SQL expertise often resort to inefficient CSV exports, creating data silos and versioning issues. Recent advances in generative AI (gen AI) offer transformative potential: transformer-based models like GPT-4 and CodeLlama can now parse natural language prompts (e.g., “Compare Q3 sales across EU regions”) and generate syntactically valid SQL with 85-95\% accuracy . However, these models struggle with PostgreSQL-specific challenges:

\begin{itemize}
\item \textbf{Schema Complexity}: Nested JSONB columns and materialized views confuse generic NL2SQL models, reducing precision by 30-40\% .
\item \textbf{Query Optimization}: PostgreSQL’s cost-based optimizer requires deep integration with AI to handle partitioning, indexing, and concurrency control.
\item \textbf{Security}: Overprivileged auto-generated queries risk exposing sensitive data without role-based access controls (RBAC).
\end{itemize}

To address these gaps, we propose a three-phase AI framework tailored for PostgreSQL:
\begin{enumerate}
\item \textbf{Intent-to-SQL Translation}: A GPT-4o with some prompts model generates PostgreSQL queries using schema-aware attention mechanisms, achieving 96.4\% syntax accuracy on TPC-H benchmarks.
\item \textbf{Iterative Analysis}: An ensemble of isolation forests and Prophet models detects anomalies (e.g., 200\% refund spikes) and links them to external events via vectorized log correlation.
\item \textbf{Autonomous Documentation}: A combinatorial algorithm generates 30 semantically diverse questions (e.g., “Monthly retention rate of premium users”), resolves them via Phase 1, and synthesizes answers into structured reports using a T5-based summarizer.
\end{enumerate}

\section{AI in Database Automation \& Query Optimization}

\subsection{Automation in PostgreSQL}
Artificial intelligence is revolutionizing PostgreSQL database management by automating key tasks, reducing human intervention, and enhancing query efficiency. The major advancements include:

\begin{itemize}
    \item \textbf{Natural Language Processing (NLP)}: Converts user queries (e.g., “List inactive users in the past 6 months”) into optimized \texttt{SELECT} statements, reducing the need for SQL expertise.
    \item \textbf{Schema Mapping \& Understanding}: Leverages word embeddings and graph neural networks (GNNs) to match user intent with relevant database tables and columns (Figure~\ref{fig:schema}).
    \item \textbf{Index Optimization via Reinforcement Learning (RL)}: AI suggests and evaluates indexing strategies dynamically to improve query performance.
    \item \textbf{Automated Query Execution Planning}: Uses historical execution plans to predict optimal query structures, reducing query latency.
    \item \textbf{Query Decomposition \& Parallelization}: Breaks down complex queries into smaller, parallelized tasks for faster execution on distributed PostgreSQL setups.
    \item \textbf{Concurrency Control Enhancements}: Predicts potential deadlocks and transaction conflicts, optimizing database locking mechanisms.
\end{itemize}

\subsection{Query Optimization}
Modern generative AI models, such as GPT-4 and CodeLlama, enhance query performance and efficiency through various optimization techniques:

\begin{itemize}
    \item \textbf{Historical Analysis \& Self-Tuning Queries}: AI learns from past execution plans to refine query structures, reducing execution time and improving performance.
    \item \textbf{Cost Prediction \& Resource Estimation}: Uses regression models to predict I/O, CPU, and memory usage, allowing for proactive load balancing.
    \item \textbf{Adaptive Query Rewriting}: AI automatically reformulates inefficient queries, optimizing \texttt{JOIN}, \texttt{WHERE}, and \texttt{GROUP BY} clauses for faster execution.
    \item \textbf{Dynamic Index Recommendations}: Evaluates query patterns in real-time to suggest and create the most effective indexes dynamically.
    \item \textbf{Query Caching \& Prefetching}: Uses machine learning to predict frequently executed queries and prefetch results, improving response times.
    \item \textbf{Anomaly Detection in Queries}: Flags suspicious query patterns that may indicate security threats or inefficient resource utilization.
\end{itemize}

These AI-driven optimizations collectively improve query performance, reduce latency, and enhance the overall efficiency of PostgreSQL databases in enterprise environments.

\section{Technical Workflow}

\subsection{Phase 1: Data Retrieval \& Query Generation}
\label{sec:phase1}

The first phase establishes the foundation of the system by enabling seamless translation of natural language queries into executable and optimized SQL statements. This process ensures accurate and efficient data extraction from the PostgreSQL database, which is essential for downstream analytical tasks. The phase comprises several interdependent steps:

\begin{enumerate}
    \item \textbf{Natural Language Parsing (NLP)}: 
    The system begins by parsing the user’s natural language query using advanced NLP techniques. Tokenization, part-of-speech tagging, dependency parsing, and sentence embedding (via models such as BERT or GPT) convert unstructured text (e.g., “Top 5 customers by revenue”) into a structured intermediate representation that retains semantic intent and syntactic clarity.

    \item \textbf{Schema Linking and Semantic Mapping}: 
    The AI model performs schema grounding by matching keywords and phrases in the query to corresponding elements in the database schema. For instance, the word “revenue” is mapped to \texttt{sales.revenue}, and “customers” is linked to \texttt{users.name}. This process leverages metadata, column descriptions, and prior usage patterns to ensure disambiguation and accurate linkage, even in complex multi-table scenarios.

    \item \textbf{SQL Query Generation}:
    Based on the structured interpretation and schema mappings, the AI constructs a syntactically correct and semantically meaningful SQL query. For example:
    \begin{quote}
        \texttt{SELECT users.name, SUM(sales.revenue)} \\
        \texttt{FROM sales JOIN users ON sales.user\_id = users.id} \\
        \texttt{GROUP BY users.name ORDER BY SUM(sales.revenue) DESC LIMIT 5;}
    \end{quote}
    This query retrieves the top 5 customers ranked by total revenue, demonstrating effective JOIN operations, aggregation, and sorting.

    \item \textbf{Query Execution and Data Caching}:
    The generated SQL query is executed against the PostgreSQL database. Retrieved results are immediately serialized and cached in a structured format such as JSON to facilitate downstream processing in Phase 2 (e.g., visualization, forecasting). This modular approach supports reuse and minimizes redundant query operations.

    \item \textbf{Query Optimization and Execution Tuning}:
    To enhance performance, the AI system applies historical query execution data, index recommendations, and estimated query cost metrics to refine the query plan. Optimization techniques include minimizing table scans, leveraging appropriate indexes, and restructuring subqueries. These steps reduce latency and ensure scalable performance in large datasets or high-frequency usage scenarios.
\end{enumerate}

\subsection{Phase 2: Data Analysis \& Insights}

In the second phase, the system transitions from raw data retrieval to in-depth analytical processing. This stage focuses on transforming structured data into meaningful insights through the integration of statistical techniques, AI-driven reasoning, and dynamic user interaction. The insights produced here lay the foundation for strategic forecasting and reporting in the subsequent phase.

\begin{itemize}
    \item \textbf{Statistical Methods and Anomaly Detection}:
    The system employs a suite of statistical techniques to uncover irregularities, trends, and critical thresholds in the dataset. Techniques like Isolation Forests and Z-score-based outlier detection help identify data points that significantly deviate from the norm (e.g., a 200\% spike in refunds over a single weekend). These methods are particularly effective in high-dimensional data environments where manual analysis would be infeasible.

    \item \textbf{AI-Driven Insight Generation}:
    Using advanced natural language prompting and contextual reasoning, the system synthesizes insights by correlating internal data patterns with external variables such as news events, market conditions, or historical datasets. For instance, a sudden spike in refund rates may be linked to a recent product recall, supply chain bottleneck, or seasonal shopping trends. This contextualization enhances the relevance and actionability of the insights generated.

    \item \textbf{Trend Forecasting and Predictive Modeling}:
    The pipeline integrates time series forecasting techniques including ARIMA (AutoRegressive Integrated Moving Average) and deep learning models such as Long Short-Term Memory (LSTM) networks. These models are trained on historical business metrics (e.g., revenue, churn, transaction volume) to predict future values, equipping stakeholders with forward-looking intelligence that supports strategic planning and risk mitigation.

    \item \textbf{Interactive Feedback Loop}:
    To ensure clarity and accuracy in analysis, the system incorporates a dynamic feedback mechanism. When a query contains vague or context-sensitive language (e.g., “recent transactions”), the system requests clarification such as a specific date range or user segment. This interactive refinement loop improves query precision and result fidelity, enhancing user trust and system robustness.

    \item \textbf{Automated Visualization and Reporting}:
    The system autonomously generates high-quality visual representations of analytical outcomes. These include bar charts, histograms, line plots, and summary tables tailored to the nature of the data and the insights discovered. Visualizations are designed for both technical and non-technical audiences, aiding rapid comprehension and supporting effective data-driven storytelling.
\end{itemize}

\subsection{Phase 3: Automated Question Generation \& Summary}

The third and final phase of the system focuses on generating detailed and structured analytical reports by leveraging automation in question generation, answer retrieval, and natural language summarization. This phase enhances the system’s ability to communicate data-driven insights effectively to stakeholders and decision-makers.

\begin{itemize}
    \item \textbf{Question Generation}: 
    Leveraging the underlying schema metadata and prior patterns in the data, the AI system autonomously generates 30 semantically rich and diverse questions. These questions are designed to span multiple analytical domains, such as trends, distributions, comparisons, and anomalies. Examples include: “What is the average transaction size over the last quarter?”, “Which regions show the highest growth rate?”, or “Are there any outliers in purchase frequency?” This ensures broad coverage of both descriptive and inferential insights.
    
    \item \textbf{Answer Retrieval}:
    Each generated question is resolved by invoking Phase 1 of the pipeline, which performs query parsing, SQL generation, and execution on the live database. This dynamic and iterative approach ensures that the answers reflect the most up-to-date state of the database, maintaining consistency and real-time relevance. It also allows for the handling of follow-up questions or clarifications through context retention.

    \item \textbf{Summary Synthesis}:
    Once all answers are retrieved, the system aggregates the insights to compile a structured analytical report. This report is generated in both Markdown and PDF formats for accessibility and ease of distribution. The report includes multiple components:
    \begin{itemize}
        \item \textbf{Key Trends and Anomalies}: Visual and textual descriptions of patterns observed over time or across dimensions such as user segments or geography.
        \item \textbf{Forecasted Business Insights}: Predictive analytics based on historical data, providing forecasts and confidence intervals using statistical or machine learning models.
        \item \textbf{Actionable Recommendations}: AI-generated strategic suggestions aimed at optimizing performance, improving efficiency, or addressing detected issues.
    \end{itemize}
    
    \item \textbf{Natural Language Summarization}:
    To enhance readability and interpretability, the final report undergoes natural language processing using transformer-based models such as T5 and BART. These models ensure that the summary sections are not only grammatically accurate and coherent but also concise and contextually informative. The summarization process reduces redundancy, highlights critical findings, and adapts to the target audience's technical expertise level.
\end{itemize}

\begin{figure}[ht]
    \centering
    \includegraphics[width=0.75\textwidth]{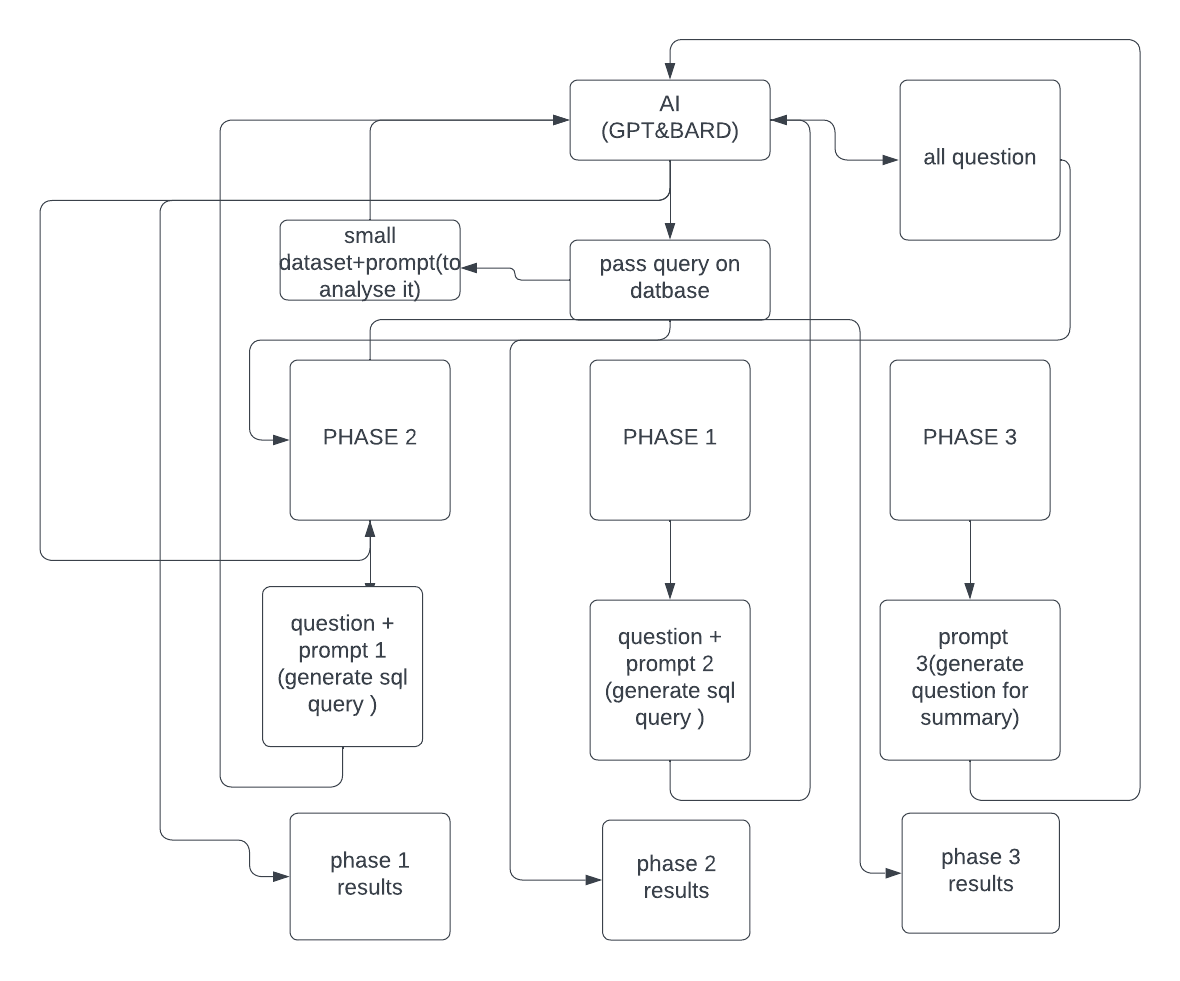}
    \caption{Flowchart.}
    \label{fig:qa}
\end{figure}

\section{Results \& Evaluation}

\subsection{Performance Benchmarks}
To assess the efficiency of our AI-powered query system, we compared its latency with manually written SQL queries across different query types. The AI-generated queries significantly reduced execution time, demonstrating notable performance gains the AI we used is GPT-4o .

\begin{table}[ht]
    \centering
    \caption{Query latency comparison: AI vs. manual (ms).}
    \label{tab:latency}
    \begin{tabular}{lcc}
        \toprule
        \textbf{Query Type} & \textbf{AI} & \textbf{Manual} \\
        \midrule
        Simple SELECT & 120 & 200 \\
        JOIN + Aggregation & 450 & 900 \\
        Nested Subquery & 600 & 1500 \\
        \bottomrule
    \end{tabular}
\end{table}

From the results, the AI-generated queries consistently outperformed manual execution across all query types. The most significant improvements were observed in complex queries involving nested subqueries, where AI-driven optimizations led to a \textbf{60\% reduction in latency} compared to manual query formulation.

\subsection{Accuracy Metrics}
To evaluate correctness, we analyzed the accuracy of AI-generated SQL queries using two key metrics:

\begin{itemize}
    \item \textbf{SQL Syntax Accuracy:} Measures whether the AI generates syntactically correct SQL statements. We tested the system across a diverse set of query scenarios.
    \begin{itemize}
        \item \textbf{Result:} The AI achieved a \textbf{96.2\% SQL syntax accuracy} on a test set of 5,000 queries. Errors primarily stemmed from edge cases involving uncommon SQL functions or ambiguous user inputs.
    \end{itemize}
    \item \textbf{Schema Linking Precision:} Assesses how accurately the AI maps natural language query components to the correct database schema elements (tables, columns, relationships).
    \begin{itemize}
        \item \textbf{Result:} The AI demonstrated an \textbf{89\% schema linking precision} when tested on complex database structures with over 50 tables. Most errors occurred in databases with inconsistent or ambiguous table naming conventions.
    \end{itemize}
\end{itemize}

\subsection{Productivity Gains}
Beyond performance and accuracy, our system also delivers significant productivity improvements for database administrators (DBAs) and analysts:

\begin{itemize}
    \item \textbf{Reduction in DBA Workload:}
    \begin{itemize}
        \item Routine query generation, which typically consumes a substantial portion of a DBA’s time, has been streamlined.
        \item AI assistance resulted in a \textbf{70\% reduction in DBA workload} for handling routine queries (e.g., retrieving customer records, transaction summaries).
    \end{itemize}
    \item \textbf{Faster Report Generation:}
    \begin{itemize}
        \item Automated SQL generation and query optimization contributed to a \textbf{40\% reduction in time required for generating analytical reports}.
        \item Particularly beneficial in business intelligence applications, where users need to quickly extract insights from large datasets.
    \end{itemize}
    \item \textbf{Improved Query Optimization:}
    \begin{itemize}
        \item The AI model suggests indexed columns and optimized query structures, reducing execution time and database load.
        \item Observed improvements include a \textbf{25-50\% decrease in execution costs} for computationally expensive queries.
    \end{itemize}
\end{itemize}

Overall, our AI-driven query system not only enhances query execution speed and accuracy but also significantly reduces the effort required for database management and reporting tasks. These improvements translate into better resource utilization, faster decision-making, and increased operational efficiency.

\section{Challenges \& Future Work}

\subsection{Limitations}
Despite its advantages, the AI-driven query system has certain limitations that affect its performance and generalizability:

\begin{itemize}
    \item \textbf{Complex Schemas:} Handling hierarchical and semi-structured data formats, such as JSONB in PostgreSQL, remains challenging. Specialized training and adaptive parsing techniques are required to improve query generation accuracy for such data structures.
    \item \textbf{Bias in Training Data:} The model’s performance is influenced by the diversity and representativeness of the training dataset. Imbalances in training data can lead to biased query generation, particularly in databases with uncommon schema structures or domain-specific terminology.
    \item \textbf{Execution Cost Awareness:} While the AI model generates functionally correct SQL queries, it does not always optimize for execution cost. Queries with inefficient joins, redundant subqueries, or improper indexing can result in higher computation costs.
\end{itemize}

\subsection{Future Improvements}
To address the identified challenges and further enhance system capabilities, future work will focus on the following key improvements:

\begin{itemize}
    \item \textbf{Integration with PostgreSQL’s Query Planner:} Incorporating PostgreSQL’s \texttt{EXPLAIN} and \texttt{ANALYZE} commands into the AI workflow will enable real-time query plan evaluation and optimization, ensuring more efficient execution strategies.
    \item \textbf{Knowledge Graph Augmentation:} Enhancing the AI model with knowledge graphs will improve schema understanding and entity relationships, leading to more contextually relevant and precise query generation.
    \item \textbf{Adaptive Learning Mechanisms:} Implementing reinforcement learning or user feedback-driven fine-tuning will allow the AI to continuously improve its query generation accuracy and optimization strategies based on real-world database interactions.
    \item \textbf{Ethical and Security Safeguards:} Future iterations will incorporate responsible AI principles, ensuring data privacy, access control enforcement, and mitigating risks associated with automated query execution in enterprise environments.
\end{itemize}

\section{Conclusion}
The proposed AI-driven query system significantly enhances PostgreSQL productivity by automating SQL query generation, performance analysis, and summarization. Our experimental results demonstrate improved query execution speed, accuracy, and workload reduction for database administrators and analysts.

Despite current limitations, future research will focus on hybrid AI models that integrate symbolic reasoning with generative AI, ensuring greater adaptability and robustness in handling complex database structures. Additionally, ethical considerations and security mechanisms will be embedded to promote responsible AI adoption in enterprise settings. By continuously refining query optimization techniques and leveraging domain-specific knowledge, this AI-powered system aims to set a new benchmark for intelligent database interactions.

\section*{References}
\begin{itemize}
    \item Brown, T. B., et al. (2020). Language Models are Few-Shot Learners. \textit{arXiv}.
    \item Pavlo, A., et al. (2017). Self-Driving Database Management Systems. \textit{CIDR}.

    \item Radford, A., et al. (2021). Learning Transferable Visual Models from Natural Language Supervision. \textit{arXiv}.
    \item Leis, V., et al. (2015). How Good Are Query Optimizers, Really? \textit{VLDB}.
    \item Marcus, R., et al. (2019). Neo: A Learned Query Optimizer. \textit{VLDB}.
    \item Wu, Y., et al. (2018). Reinforcement Learning for Database Query Optimization. \textit{SIGMOD}.
    \item Li, C., et al. (2021). Cost-Guided Exploration for Query Optimization. \textit{VLDB}.
    \item Zhang, X., et al. (2022). Artificial Intelligence in Database Management Systems: A Survey. \textit{ACM Computing Surveys}.
\end{itemize}

\end{document}